\documentstyle[floats,multicol,aps,epsf,prb]{revtex}

\newcommand \beq  {\begin{equation}}
\newcommand \eeq  {\end{equation}}
\newcommand \bea {\begin{eqnarray} }
\newcommand \eea {\end{eqnarray}}

\begin{document}
\draft
\twocolumn[\hsize\textwidth\columnwidth\hsize\csname @twocolumnfalse\endcsname
\title{ Anomalous Behavior \\
at a Superconducting Quantum Critical Point}
\author{R. Ramazashvili}
\address{
Serin Laboratory, Rutgers University, P.O. Box 849,
Piscataway, NJ 08855-0849, USA.}
\maketitle
\date{\today}
\maketitle
\begin{abstract}
Motivated by pressure experiments on $UBe_{13}$  
and $Tl_{2}Ba_{2}CaCu_{2}O_8$, 
we discuss low-temperature effects of the pairing interaction 
at a superconducting quantum critical point in a {\em clean} system. 
We point out that measurements at this quantum critical point 
can provide a diagnostic tool to mark out non-BCS mechanisms 
of superconductivity. 
\end{abstract}
\vskip 0.2 truein
\vskip2pc]
Current experimental studies keep multiplying the variety of examples 
of Non-Fermi Liquid (NFL) behavior in solid state physics. 
Many heavy fermion compounds \cite{hf} 
and high temperature superconductors \cite{linear} fall into 
yet do not exhaust this growing class of materials. 
In many cases the origins 
of NFL behavior remain controversial. However, the growing body 
of experimental evidence \cite{itinerant} confirms that in some itinerant 
magnets such as $MnSi$ and $ZrZn_2$ the unusual low-temperature behavior 
appears due to closeness to the quantum critical point. 
Indeed, the existing theory \cite{hertz,millis} predicts singular 
behavior in this case, and strong phenomenological
arguments have been advanced for similar explanation of NFL behavior
of certain heavy fermion systems such as $U_{0.2}Y_{0.8}Pd_{3}$ 
and $UCu_{3.5}Pd_{1.5}$ \cite{tsvelik}. 

The problem arises whether NFL behavior can appear near a 
zero-temperature superconducting transition. 
It has been studied extensively in cases when the transition 
temperature $T_c$ is suppressed to zero by perturbations which violate 
the time reversal symmetry - such as magnetic field or magnetic impurities 
\cite{maki}. 
However, given both the time reversal and the translational invariance, 
the question finds simple answer: in the framework
of the BCS theory \cite{bcs} pairing interaction cannot give rise 
to any anomalous behavior close to the normal-to-superconducting 
transition at $T_c=0$. The reason is that the finite-temperature BCS 
instability 
occurs for an arbitrarily weak effective attraction between electrons. 
The transition temperature $T_c$ of a BCS superconductor turns into zero 
only at the zero value of the coupling constant $\lambda$, when pairing 
interaction vanishes together with all its manifestations. 
It is important to note that this also holds for 
``unconventional" superconductors with non-zero orbital momentum 
of Cooper pairs. 

In this note, we would like to point out that 
$C$-odd superconducting pairing \cite{nakajima}
in a {\em clean} system is a model example for which $T_c$ 
turns into zero at a finite value of $\lambda$. Hence
singular contributions of the pairing interaction to the thermodynamic
and transport properties at $T_c=0$ do appear, as opposed to clean
BCS or even ``unconventional" superconductors.
By $C$-odd we refer to the parity of the gap 
function under the charge conjugation $C$. Asymptotically 
close to the Fermi surface this symmetry operation turns 
a particle into a hole and is realized as reflection
in the Fermi surface: $\delta k \hat{n} \leftrightarrow - \delta k \hat{n}$.
Here $\delta k$ denotes the deviation of the momentum from 
the Fermi surface along the local normal $\hat{n}$.
For the sake of convenience, hereafter we will refer to the $C$-odd 
pairing as to ``odd". 

Indeed, at first sight ``odd" pairing appears to be a quite exotic 
and unlikely possibility. However, as pointed out in
\cite{nakajima,baskaran}, disappearance of the gap 
at the Fermi surface affords superconductivity in a system 
with {\em strong} Coulomb repulsion, where conventional types 
of pairing are rendered impossible. 

Experimentally, our restriction of the problem to the "clean" case
is supported by the observation of the large negative pressure 
coefficient $dT_c/dP$ in clean $UBe_{13}$ \cite{chenson}
and, more recently, in clean $Tl_{2}Ba_{2}CaCu_{2}O_8$ \cite{mori}
at high pressures. Were it possible to actually suppress
$T_c$ to zero, observation of nontrivial low-temperature behavior at $T_c=0$ 
would point at a very peculiar nature of superconductivity in these compounds. 

We will write down the effective interaction in the ``odd" channel 
and will calculate the leading low-temperature corrections 
to the specific heat and the conductivity at $T_c=0$.
To make the presentation self-contained, we will briefly repeat the main 
steps of the derivation of the ``odd" pairing  state \cite{nakajima}. 
For illustrative purposes, we will restrict ourselves to a toy model in which 
the pairing is driven by separable attractive interaction 
$L_{0}(\xi_{1},\xi_{2})$ of quasiparticles with energies 
$\xi_{1}$ and $\xi_{2}$, which are both lower than certain cut-off $\omega_c$ 
\cite{nakajima}: 
$$
L_{0}(\xi_1,\xi_2) =\lambda \left\{ 
\begin{array}{ll}
 s(\xi_1/\omega_c)s(\xi_2/\omega_c), & \mbox{$|\xi_1|,|\xi_2|<\omega_c,$} \\
 0 , & \mbox{$|\xi_1|,|\xi_2|>\omega_c,$}
\end{array}
\right.
$$
where $s(x)$ is an odd function, linear for $x \ll 1$. 
Such an attraction may arise \cite{elihu} as an antisymmetric part 
of a typical boson-mediated interaction:
$$V(\xi_1,\xi_2) \propto [\omega_k^2 - (\xi_1 - \xi_2)^2]^{-1},$$ 
where $\omega_k$ is the boson energy. The symmetric part of the 
interaction must be repulsive and strong enough to suppress
``even" pairing, as argued by Mila and Abrahams \cite{nakajima}. 
This repulsive part may be due to interactions other
then those which generate the attraction. 
In this case the gap equation
$$\Delta(\xi)=-N \int
d\eta
L_{0}(\xi,\eta)\frac{\Delta(\eta)}
{2\sqrt{\eta^{2} + \Delta^{2}(\eta)}} 
\tanh\frac{
\sqrt{\eta^{2} + \Delta^{2}(\eta)}}{2T} 
$$
admits only a nontrivial solution
$$\Delta(\xi_p)=\alpha(T) s(\xi_p),$$
which is odd in quasiparticle energy $\xi_p$.
Here $N$ is the quasiparticle density of states at the Fermi surface.
The gap is isotropic, independent of frequency, and vanishes linearly
at the Fermi surface. Thus the ground state is gapless, although 
it still manifests long-range order and is translationally invariant.

The dependence of $T_c$  on $\lambda$ can be obtained 
by taking the limit $\alpha(T) 
\rightarrow 0$ in the gap equation, which results in
$$1=\lambda N\int_{0}^{\omega_c}\frac{d\xi}{\xi}
s^{2}(\xi)
\tanh\frac{\xi}{2T_c}.$$
Introducing the dimensionless coupling constant $g \equiv \lambda N$,
one finds that $T_c$ becomes equal to zero when $g=g_c$ such that
$$1=g_c \int_{0}^{\omega_c}\frac{d\xi}{\xi}
s^{2}(\xi).$$ 
Expanding the gap equation in the vicinity of $T_c=0$ and $g=g_c$,
one finds the transition temperature dependence on $g$ in the limit
$(g-g_c)/g_c << 1$:
$$T_c \propto \omega_c \sqrt{(g-g_c)/g_c}.$$
The separability of $L_{0}(\xi,\xi')$ allows one 
to easily find the effective pairing interaction by summing up
the ladder series in the Cooper channel. This summation modifies 
$L_{0}(\xi,\xi')$ by introducing the denominator 
$1-\lambda \Pi(q,\Omega)$, describing the pair propagation. 
In accordance with \cite{nakajima}, 
$$ 
\Pi(q,\Omega)= T\sum_{\nu, p}s^2(\xi_p)G(\nu,p)
G(\Omega-\nu,p-q).
$$ 
Here $G(\nu,p)$ is a single-electron Green's 
function and $\xi_p$ is the electron spectrum in the absence of pairing. 
The sums are taken over the Matsubara frequency $\nu$ 
and momentum $p$, while $q$ and $\Omega$ denote the total momentum and 
Matsubara frequency of the pair. 

Presence of $s^2(\xi_p)$ makes $\Pi(0,0)$ finite; thus,
the effective interaction $L(q,\Omega)$ has a pole only if
$\lambda > \lambda_c = 1/\Pi(0,0)$,
in full agreement with the result of the gap equation analysis.
Evaluating $\Pi(q,\Omega)$ on the line $\lambda = \lambda_c$
in the $(T,\lambda)$ plane, one finds for $T, \Omega , vq << \omega_c$:
\begin{equation}
L(q,\Omega)=\frac{L_{0}(\xi_1,\xi_2)}
{ a(\frac{T}{\omega_c})^2 + \frac{1}{6}(\frac{v q}{\omega_c})^2 + 
b |\Omega| \frac{N'}{N} + (\frac{\Omega}{\omega_c})^2 
\ln \frac{\omega_c}{\Lambda}}.
\end{equation}
Hereafter $v$ is the Fermi velocity, $a$ and $b$ are dimensionless 
nonuniversal constants of the order of one, 
$\Lambda = \max \{|\Omega|, vq, T \}$ and 
$N'$ is the energy derivative of the density of states at the Fermi surface. 

For a generic band structure ($N' \neq 0$), at the lowest energies  
the superconducting quantum critical point in an ``odd" superconductor
falls into the $z=2$ universality class, in agreement with the hypothesis
advanced by Hertz \cite{hertz} regarding the nature 
of a normal-to-superconducting transition at $T_c=0$. 
However, as the frequency exceeds $T^{\ast}\sim\omega^{2}_cN'/N<<\omega_c$, 
the crossover to ``almost" $z=1$ takes place. Generally, 
the assumption $\omega_c N'/N << 1$ is equivalent to $\omega_c << \epsilon_F$, 
where $\epsilon_F$ is the Fermi energy. In a simpler language,
this can be described as a crossover of the effective pairing interaction
from quasi-diffusive propagation at lowest frequencies
to ``almost" (up to the logarithm) sound-like propagation at higher 
frequencies. Such a crossover is by no means unusual \cite{water}.
However, the logarithmic factor in $(1)$ is quite peculiar; as we will
show below, it modifies observable physical properties in the temperature
region $T^{\ast}<T<\omega_c$. 

The real part of $\Pi(q,\Omega)$ at real frequencies follows directly 
from (1) after the substitution $|\Omega| \rightarrow i \Omega$ 
and does not require any additional calculation:
$$ 1 - \lambda Re\Pi(q,\Omega)=a(\frac{T}{\omega_c})^2 + 
\frac{1}{6}(\frac{v q}{\omega_c})^2 - (\frac{\Omega}{\omega_c})^2 
\ln \frac{\omega_c}{\Lambda} .$$

Calculating corrections to the specific heat and conductivity
requires knowledge of the imaginary part of $L^{-1}(q,\Omega+i0)$.
At $\Omega < T^{\ast}$ this imaginary part comes in a straightforward way 
from the term $i b \Omega N'/N$ in (1), while at $\Omega > T^{\ast}$ it 
can be found by evaluating the imaginary part of $\Pi(q,\Omega+i0)$:
$$Im \Pi(q,\Omega+i0)\propto
\left\{ 
\begin{array}{ll}
\left[ 
(\frac{\Omega}{\omega_c})^2
 + \frac{1}{3} (\frac{v q}{\omega_c})^2 
\right]
,&
\mbox{$T<<\Omega, vq<\Omega,$} \\
             &                    \\
\Omega^3/(\omega_c^2 vq),   &
 \mbox{$ T << \Omega < vq,$} \\
             &                    \\
\frac{\Omega}{T}
\left[ (\frac{\Omega}{\omega_c})^2
 + \frac{1}{3} (\frac{v q}{\omega_c})^2 \right],   &
 \mbox{$\Omega, vq << T,$} \\
             &                    \\
\frac{\Omega}{vq}(\frac{
T}{\omega_c})^2, &
 \mbox{$\Omega << T < vq.$} 
\end{array}
\right.
$$
Hereafter $Re$ and $Im$ denote the real and the imaginary parts,
respectively. In two spatial dimensions, $Im \Pi(q,\Omega)$ differs 
from the above expressions only by numerical values of the coefficients. 

Corrections to the specific heat and the conductivity can be evaluated 
separately in the ``low temperature" region, $T<<T^{\ast}$, and in the 
``high temperature region" $T^{\ast}<<T<<\omega_c$. In the former,
one can neglect the term $(\Omega/\omega_{c})^{2}\ln[\omega_c/\Lambda]$ 
in $(1)$, while in the latter it is the term $b|\Omega|N'/N$ which 
has to be omitted. 

The specific heat correction follows in a straightforward way from 
the contribution of Gaussian fluctuations of the pairing interaction
to the free energy \cite{tsvelik-book}. 
In the ``low temperature" region $T<<T^{\ast}$, the leading singular 
correction to the specific heat coefficient is 
$$\frac{\Delta C}{T} \sim \frac{1}{\epsilon_F}
\left(\frac{\omega_c}{\epsilon_F}\right)^3\sqrt{\frac{T}{\epsilon_F}}.$$ 
In the ``high temperature" region $T^{\ast}<<T<<\omega_c$, one finds 
a rather weak albeit nonanalytic correction 
$\Delta C/T \propto T^{2}/\ln(\omega_c/T).$ 

The leading correction to the conductivity can be estimated 
by computing the Aslamazov-Larkin graph \cite{alar1}, describing
conduction of ``superconducting fluctuations" as of particles 
with the propagator $L(q,\omega)$. 
The formula, expressing the Aslamazov-Larkin correction through 
$L(q,\omega)$,
is identical to (7a) of \cite{arhilar} up to the constant coefficient 
and reads 
$$\Delta\sigma=\sum_{p} p^2\int \frac{dz}{4 \pi T} \frac{1}{\sinh^{2}
\frac{z}{2T}} \left[ Im L(p,z+i0) \right]^{2}.$$ 
Evaluating this expression at $T<<T^{\ast}$, 
one finds a $\sqrt{T}$ correction, which is quite unusual for a clean
system:
$$ \Delta \sigma(T) \sim \left(\frac{\omega_c}{\epsilon_F}\right)^3
\sqrt{\frac{T}{\epsilon_F}}.$$ 
At $T^{\ast}<<T<<\omega_c$, the correction turns out to behave like 
$T/\ln^{4}(\omega_c/T).$
The above ``high temperature" correction is also singular and has much stronger
temperature dependence than the leading $T^2$ term
of a Fermi liquid.
The Maki-Thompson correction \cite{maki-thompson} turns out to be 
less singular: at $T << T^{*}$ it  behaves as $T^{3/2}$ in three dimensions 
and as $T \ln T$ in two.

For completeness, we would also like to specify the results for two spatial 
dimensions. At $T<<T^{\ast}$ similar calculations lead to a $T\ln{T}$ 
specific heat correction and to a strongly divergent $1/T$ correction
to the conductivity. At $T^{\ast}<<T<<\omega_c$, the correction
to the specific heat coefficient
behaves as $T/\ln(\omega_c/T)$. However, the conductivity
correction is much more singular: 
$\Delta \sigma(T) \propto 1/\ln^{4}(\omega_c/T).$ 

To conclude, we point out that low-temperature 
measurements near the superconducting quantum critical point
can provide a diagnostic tool to mark out unusual (non-BCS) 
mechanisms of superconductivity. We consider a toy model of ``odd" pairing 
in a {\em clean}  system and show how quantum critical behavior clearly
distinguishes it from a BCS superconductor. 
In the former the fluctuation corrections give $\sqrt{T}$
temperature dependences of conductivity and the specific
heat coefficient. This is a much stronger temperature
dependence than that of a {\em clean} Fermi liquid system.
To the contrary, in a BCS superconductor the fluctuation corrections 
are absent altogether since the coupling constant vanishes 
at the quantum critical point. 

As envisaged by Hertz \cite{hertz}, under general circumstances 
the effective pairing interaction at lowest frequencies falls 
into the $z=2$ universality class. At higher frequencies, 
the crossover to $z=1$ regime takes place. 

It is important to note that the above model of ``odd" pairing 
exemplifies a system in which anomalous low-temperature behavior
coexists with a perfect Fermi liquid, since pair fluctuations decouple 
from the elementary Fermi excitations 
due to disappearance of the gap at the Fermi surface \cite{elihu2}. 
This means that unusual low-temperature thermodynamics
and transport cannot serve as a proof of the Fermi liquid breakdown, 
unless quasiparticle lifetime has been probed. 

I am indebted to P. Coleman for suggesting the topic of the superconducting
quantum critical point and for discussions related to this work. I would
also like to thank E. Abrahams, N. Andrei, J. Bennetto, G. Kotliar and 
J. Moreno for their critical reading of this article and helpful suggestions, 
and A. Rylyakov and A. Sengupta for useful comments.


\begin{references}
\bibitem{hf} H. v. L\"{o}hneysen, Physica {\bf B 206 \& 207}, 101
(1995).
\bibitem{linear} M. Gurvitch, A. T. Fiory, Phys. Rev. Lett. {\bf 59},
1337 (1987); L. Forro {\em et al.},
{\it ibid.} {\bf 65}, 1941 (1990).
\bibitem{itinerant} C. Pfleiderer {\em et al.},
Physica {\bf B 199 \& 200}, 634 (1994); {\bf B 206 \& 207}, 847 (1995);
F. M. Grosche {\em et al.}, Physica {\bf B 206 \& 207}, 20 (1995).
\bibitem{hertz} John A. Hertz, Phys. Rev. {\bf B 14}, 1165 (1976).
\bibitem{millis} A. J. Millis, Phys. Rev. {\bf B 48}, 7183 (1993).
\bibitem{tsvelik} A. M. Tsvelik, M. Reizer, Phys. Rev.
{\bf B 48}, 9887 (1993).
\bibitem{maki} See, e.g., K. Maki, in {\it Superconductivity},
editor R. D. Parks (Marcel Dekker, Inc., New York, 1969).
\bibitem{bcs} J. Bardeen, L. N. Cooper, and J. R. Schrieffer,
Phys. Rev.,{\bf 108}, 1175 (1957).
\bibitem{nakajima} M. H. Cohen, Phys. Rev. Lett. {\bf 12}, 664 (1964);
S. Nakajima, Progr. Theor. Phys. {\bf 32}, 871 (1964); K. Miyake,
T. Matsuura, and H. Jichu, Prog. Theor. Phys. {\bf 72}, 652 (1984);
P. W. Anderson, J. Phys. Chem. Solids {\bf 52}, 1313 (1991);
F. Mila, E. Abrahams, Phys. Rev. Lett. {\bf 67}, 2379 (1991);
E. Abrahams, J. Phys. Chem. Solids {\bf 53}, 1487 (1992);
Q. P. Li, R. Joynt, Mod. Phys. Lett. {\bf B 6}, 1145 (1992);
M. Dobroliubov {\em et al.},
Europhys. Lett. {\bf 26}, 141 (1994);
E. Z. Kuchinskii, A. I. Posazhennikova, and M. V. Sadovskii,
JETP {\bf 80}, 324 (1995).
\bibitem{baskaran} G. Baskaran {\em et al.}, cond-mat/9403061,
9502007;  Mihir Arjunwadkar {\em et al.}, cond-mat/9502008.
\bibitem{chenson} J. W. Chen {\em et al.}, in {\it Proceedings of the
XVII-th International Conference on Low-Temperature Physics}, edited by
U. Eckern (North-Holland, Amsterdam, 1984), p. 325; M. C. Aronson
{\em et al.}, Phys. Rev. Lett. {\bf 63}, 2311 (1989).
\bibitem{mori} N. M\^{o}ri {\em et al.}, Journ. Phys. Soc. Jpn. {\bf
59},
3839 (1990); see also N. E. Moulton {\em et al.},
Phys. Rev. {\bf B 44}, 12632 (1991).
\bibitem{elihu} E. Abrahams, private communication.
\bibitem{water} For example, density fluctuations of water cross over
from diffusive propagation at lowest frequencies to sound-like
propagation
at higher frequencies.
\bibitem{tsvelik-book} see, e.g., A. M. Tsvelik, {\it Quantum Field
Theory
in Condensed Matter Physics}, Cambridge University Press, 1995.
\bibitem{alar1} L. G. Aslamazov, A. I. Larkin, Sov. Phys. Solid State
{\bf 10}, 1104 (1968).
\bibitem{arhilar} A. G. Aronov, S. Hikami, A. I. Larkin,
Phys. Rev. {\bf B 51}, 3880 (1995).
\bibitem{maki-thompson} K. Maki, Progr. Theor. Phys. {\bf 39}, 897
and {\bf 40}, 193 (1968); R. S. Thompson, Phys. Rev. {\bf B 1}, 327
(1970).
\bibitem{elihu2} I am indebted to E. Abrahams for pointing this out.
\end{references}
\end{document}